\documentclass[%
 reprint,
 amsmath,amssymb,
 aps,
 pra,
]{revtex4-1}

\usepackage{graphicx}
\usepackage{dcolumn}
\usepackage{bm}
\usepackage{braket}
\usepackage{color}

\makeatletter
\def\authornote{\xdef\@thefnmark{$\dagger$}\@footnotetext}
\makeatother

\begin{document}

\preprint{APS/123-QED}

\title{Pulsed coherent drive in the Jaynes--Cummings model}
\author{Kevin Fischer$^\dagger$}
 \email{kevinf@stanford.edu}
\author{Shuo Sun$^\dagger$}%
 \authornote{These authors contributed equally.}
\author{Daniil Lukin}%
\author{Yousif Kelaita}%
\author{Rahul Trivedi}%
\author{Jelena Vu\v{c}kovi\'c}%
\affiliation{E. L. Ginzton Laboratory, Stanford University, Stanford CA 94305, USA}%

\date{\today}

\begin{abstract}
The Jaynes--Cummings system is one of the most fundamental models of how light and matter interact. When driving the system with a coherent state (e.g. laser light), it is often assumed that whether the light couples through the cavity or atom plays an important role in determining the dynamics of the system and its emitted field. Here, we prove that the dynamics are identical in either case except for the offset of a trivial coherent state. In particular, our formalism allows for both steady-state and the treatment of any arbitrary multimode coherent state driving the system. Finally, the offset coherent state can be interferometrically canceled by appropriately homodyning the emitted light, which is especially important for nanocavity quantum electrodynamics where driving the atom is much more difficult than driving the cavity.
\end{abstract}

\maketitle

\begin{figure}[b]
  \includegraphics{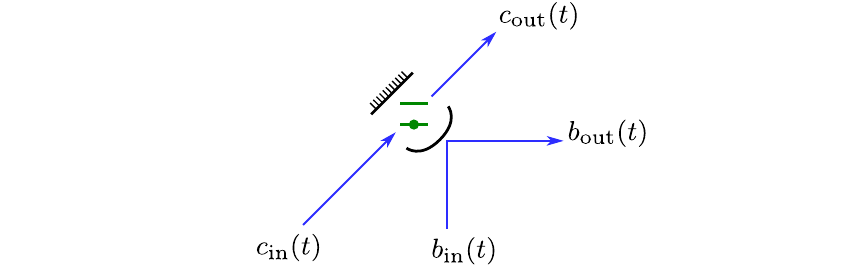}
  \caption{Schematic of a Jaynes--Cummings system, containing a two-level quantum system coherently interacting with a cavity mode. The cavity couples to the reservoir $b$, while the two-level atom couples to reservoir $c$.}
  \label{fig:0}
\end{figure}

In this work, we study the Jaynes--Cummings (JC) Hamiltonian driven by a coherent state. The JC system consists of a single bosonic mode (e.g. an electromagnetic cavity) coherently interacting with a single fermionic mode (e.g. an electron or two-level atom). These modes are radiatively coupled to external reservoirs, which can cause the system to both leak or absorb energy depending on their state. Preparing the reservoirs with coherent light is one of the most common ways of pumping energy into the system, and provides an excellent way to understand its quantum-mechanical responses. Despite the simplicity of the JC model itself (Fig. 1), it gives rise to a variety of complex phenomena when driven. For example,  vacuum Rabi splitting \cite{khitrova2006vacuum,ota2015vacuum}, photon blockade \cite{muller2015coherent,hamsen2017two,tang2015quantum} and tunneling \cite{dory2017tuning}, bistability \cite{bishop2010response, peano2010quasienergy} or symmetry breaking \cite{alsing1991spontaneous}, squeezed states \cite{alsing1992dynamic}, Mollow triplets \cite{fischer2016self} or state dressing \cite{hopfmann2017transition}, and a rich structure of multi-photon resonances \cite{laussy2012climbing,munoz2018filtering} have all been observed, as well as finding use in the readout of qubits \cite{reed2010high,sun2017cavity}. There are multiple ways to drive the JC system, as shown in Fig. 1. In particular, the coherent state could be prepared in the reservoirs coupled to the cavity or coupled to the atom \cite{radulaski2017nonclassical}. Almost all of the above studies have shown substantive differences in observed radiation from the JC system depending on the driving configuration. With many of these works, observing the differences between reservoir initialization is a primary portion of the result.

To understand the perceived importance of which reservoir contains the coherent state, consider the JC level structure (Fig. 2). The two-level atom coherently interacts with the cavity at rate $g$, breaking the harmonic ladder of uncoupled levels. The new eigenstates of the system are called polaritons because they contain equal superpositions of light and matter, and are indexed by the labels $\ket{\pm,n}$. The anharmonicity of these eigenstates causes the JC system to have a highly nonlinear response to input light, leading to its wide variety of quantum effects. However, the magnitude of the nonlinearity experienced by light driving the system through the cavity or atom appears to be different in this picture. Notably, the allowed transitions through the ladder are different depending on the reservoir coupling. Allowed transitions through the cavity keep the same polariton symmetry $\ket{\pm,n}\leftrightarrow\ket{\pm,n+1}$ while the allowed transitions through the atom alternate $\ket{\pm,n}\leftrightarrow\ket{\mp,n+1}$. The anharmonicity between successive eigenstates of the same symmetry is lower than for successive eigenstates of alternating symmetry. Consequently, the effective nonlinearity in this picture is higher for light driving the atom than the cavity. From this perspective, one might easily imagine that driving the atom is a far more efficient way to obtain a strong nonlinearity than driving the cavity.

Experimentally, it is often far easier to drive the cavity than the atom, which posed a major impediment to observing a strong nonlinear response in early JC systems based on solid-state platforms (e.g. see \citet{faraon2008coherent} or \citet{reinhard2012strongly}). From our experimental and theoretical work \cite{fischer2016self,muller2016self,fischer2017chip,sun2017cavity}, we now understand that despite driving the cavity, it is actually possible to recover an atom-like driving response by using homodyne interference.

In the steady state, it is known that the dynamics under drive through the cavity or atom can be related via a trivial transformation \cite{alsing1991spontaneous,alsing1992dynamic,carmichael2009statistical}. For this work, we extend the analysis to an arbitrary input coherent pulse and demonstrate there is a simple transformation between driving with the cavity or atom in the general case. Specifically, we show that driving the cavity can be seen simply as building up a coherent state in the cavity as if it were empty, and that coherent state drives the atom through the atom-cavity coupling. This result is potentially important for quantum information processing, where cavity-atom systems typically manipulate pulses of laser light rather than steady-state quantities \cite{o2009photonic}.

\begin{figure}
  \includegraphics{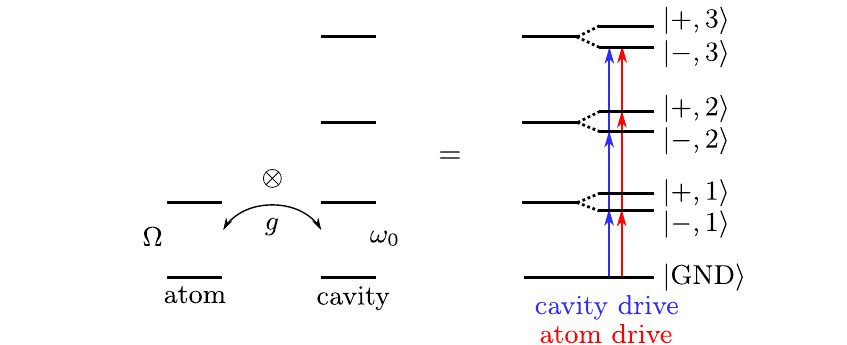}
  \caption{Energy-level structure of a Jaynes--Cummings system when the atom and cavity have the same natural oscillation frequency. Blue arrows denote allowed transitions from the reservoir-cavity coupling, while red arrows denote allowed transitions from the reservoir-atom coupling.}
  \label{fig:1}
\end{figure}

To begin, we discuss a mathematical model for the Jaynes--Cummings system. The model describes a single cavity mode and a single two-level atom, which coherently exchange energy at the rate $g$. The resulting Hamiltonian is given by
\begin{eqnarray}
H=\omega_0 a^\dag a +\Omega\sigma^\dag \sigma + \text{i}g\left(a^\dag \sigma-\sigma^\dag a\right),
\end{eqnarray}
where $\omega_0$ and $\Omega$ are the natural frequencies of the cavity mode $a$ and two-level atom $\sigma$, respectively. The bosonic cavity obeys $[a,a^\dag]=1$, while the fermionic atom obeys $\{\sigma,\sigma^\dag\}=1$. For simplicity, we consider that the cavity and atom couple to independent reservoirs, $b$ and $c$, respectively (Fig. 1). By having these two reservoirs coupled to the Jaynes--Cumming system, we implicitly constrain our analysis to an open-quantum systems perspective.

The cavity couples to the photonic reservoir $b$ through the operator $a$ at a rate $\kappa$, while the atom couples to the reservoir $c$ through $\sigma$ at rate $\gamma$. This coupling leads to spontaneous emission, but also allows for energy to be injected into the system from a laser pulse. These processes can be described by the Heisenberg--Langevin equations \cite{gardiner2004quantum} for an arbitrary system operator $x$
\begin{eqnarray}
\dot{x}&=&-\text{i}\left[x, H\right]\nonumber\\
&&-\left[x,a^\dag\right]\left(\frac{\kappa}{2}a+\sqrt{\kappa}b_\text{in}(t)\right)+\left(\frac{\kappa}{2}a^\dag+\sqrt{\kappa}b^\dag_\text{in}(t)\right)\left[x, a\right]\nonumber\\
&&-\left[x,\sigma^\dag\right]\left(\frac{\gamma}{2}\sigma+\sqrt{\gamma}c_\text{in}(t)\right)+\left(\frac{\gamma}{2}\sigma^\dag+\sqrt{\gamma}c^\dag_\text{in}(t)\right)\left[x, \sigma\right],\nonumber\\\label{eq:HL}
\end{eqnarray}
where the input and output operators are defined for reservoir mode $o$ ($b$ or $c$ here) as
\begin{subequations}
\begin{eqnarray}
o_\text{in}(t)=\frac{1}{\sqrt{2\pi}}\int\mathop{\text{d}\omega}\text{e}^{-\text{i}\omega(t-t_0)}o_0(\omega) \\
o_\text{out}(t)=\frac{1}{\sqrt{2\pi}}\int\mathop{\text{d}\omega}\text{e}^{-\text{i}\omega(t-t_1)}o_1(\omega).
\end{eqnarray}
\end{subequations}
The operators $o_0(\omega)$ and $o_1(\omega)$ are the Heisenberg operators $o(\omega)$ at times $t_0$ and $t_1$, respectively. Consequently, the input and output modes obey the relations $[o_\text{in}(t), o_\text{in}^\dagger(s)]=\delta(t-s)$ and $[o_\text{out}(t), o_\text{out}^\dagger(s)]=\delta(t-s)$. Further, operators between different reservoir modes commute. The input modes represent the Fourier transformed fields at time $t_0$ \textit{forward} propagated to time $t$, while the output modes represent the transformed fields at time $t_1$ \textit{backward} propagated to time $t$.

Applying Eq. \ref{eq:HL} for the Jaynes--Cummings system yields a closed set of Langevin equations
\begin{subequations}\label{eq:langevin}
\begin{align}
&\dot{a}=-\left(\text{i}\omega_0+\frac{\kappa}{2}\right)a+g\sigma-\sqrt{\kappa}b_\text{in}\label{eq:a}\\
&\dot{\sigma}=-\left(\text{i}\Omega+\frac{\gamma}{2}\right)\sigma+\sigma_z\big(ga+\sqrt{\gamma}c_\text{in}(t)\big)\label{eq:sigma}\\
&\dot{N}=-\gamma N-\sigma^\dag\big(ga+\sqrt{\gamma} c_\text{in}(t)\big)-\big(ga^\dagger+\sqrt{\gamma}c_\text{in}^\dag(t) \big)\sigma,\label{eq:N}
\end{align}
\end{subequations}
where $\sigma_z=\sigma^\dag\sigma-\sigma\sigma^\dag$ and $N=(\sigma_z+1)/{2}$. The equation of motion \ref{eq:a} for the cavity mode is linear, but the two-level system can only store a finite amount of energy, resulting in the nonlinear Eqs. \ref{eq:sigma}-\ref{eq:N}.

Lastly, there is an important set of boundary conditions between the input and output fields
\begin{subequations}\label{eq:inout}
\begin{eqnarray}
b_\text{out}(t)&=&b_\text{in}(t)+\sqrt{\kappa}a(t)\\
c_\text{out}(t)&=&c_\text{in}(t)+\sqrt{\gamma}\sigma(t),
\end{eqnarray}
\end{subequations}
which explains the output field in a reservoir is the linear combination of the input field and the radiated field into that channel. There also exists a causality relationship between the system operators and the input/output fields at different times, however we will not make use of those commutators in this work.

We will next explore the effects on adding input coherent states to the modes $b_\text{in}$ or $c_\text{in}$. Our goal is to show that the expectation values of any Heisenberg operators of the system are governed by the same dynamics, regardless of whether driven by the cavity or atom. For any Heisenberg-picture operator $S$ involving the output fields, its expectation value is given by $\braket{\Psi_0 |S| \Psi_0}$, where $\ket{\Psi_0}=\ket{\phi}\otimes \ket{\psi}$ is the initial state of the JC system $\ket{\phi}$ and reservoirs $\ket{\psi}$.

Now, we consider a specific type of reservoir state which can be written as a unitary transformation applied on vacuum $\ket{\psi}=U\ket{\textrm{vac}}$. Then, the expectation of different input or output operators $s(t)$ (from any of the photonic reservoirs) is
\begin{eqnarray}
\braket{\Psi_0|U^\dag s_1(t_1)s_2(t_2)\ldots s_n(t_n)U|\Psi_0}=\hspace{20pt}\nonumber\\
\braket{\Psi_0| \tilde{s}_1(t_1)\tilde{s}_2(t_2)\ldots \tilde{s}_n(t_n)|\Psi_0}.\label{eq:mollow}
\end{eqnarray}
This relation follows from $\tilde{s}(t)=U^\dag s(t) U$ and $U^\dag U=1$. In the case where we let $U$ be a displacement operator $D$ that creates coherent fields in the input photonic modes, then this transformation is called the Mollow transformation. Hence, it is only necessary to find each unique transformed operator in order to fully specify the output field of the system.

To be more concrete, explicitly let $U$ be some combination of displacement operators
\begin{eqnarray}
D_o[\delta]=\exp\left(\int\mathop{\text{d} t} \big\{\delta(t)o_\text{in}^\dag(t) - \delta^*(t)o_\text{in}(t) \big\}\right),
\end{eqnarray}
where $D_o[\delta]$ creates the coherent state $\delta$ in the photonic reservoir $o$, which could potentially represent $b$ or $c$ in our model. Hence, the displacement operators commute with every other input field and Schr\"{o}dinger-picture system operator. The action of the Mollow transformation on input operators is given by
\begin{eqnarray}
D_o^\dag[\delta] o_\text{in}(t) D_o[\delta] = o_\text{in}(t) + \delta(t).
\end{eqnarray}
Consequently, the effect of a Mollow transformation on the total system dynamics can be understood as follows. For all operators, the transformation would yield identical dynamics as Eqs. \ref{eq:langevin}-\ref{eq:inout}, except the input operators need to be modified ${o}_\text{in}(t) \rightarrow o_\text{in}(t) + \delta(t)$ to account for all potential input fields~$\{\delta\}$.

Now, consider two specific cases for driving the Jaynes--Cummings system that correspond to relevant experimental situations:
\begin{enumerate}
\item Only the emitter is driven by a coherent state $\zeta$
\begin{eqnarray}
U=D_c[\zeta].
\end{eqnarray}
Then, in the Mollow transformation all of the system operators and output fields obey equivalent dynamics as Eqs. \ref{eq:langevin}-\ref{eq:inout} except
\begin{subequations}
\begin{align}
&\dot{{\sigma}}\rightarrow\dot{\sigma}-\sqrt{\gamma}\sigma_z\zeta(t)\\
&\dot{{N}}\rightarrow\dot{N}+\sqrt{\gamma}\sigma^\dag \zeta(t) + \sqrt{\gamma}\zeta^*(t)\sigma\\
&{c}_\text{out}(t)\rightarrow c_\text{out}(t) - \zeta(t).
\end{align}
\end{subequations}
This corresponds to the situation where the laser pulse in mode $c$ comes in and imparts energy to the emitter, and causes the system's dynamics to be imprinted onto the output fields. After interaction, the input pulse continues traveling along mode $c$ and is found in the output channel ${c}_\text{out}$.

\item Only the cavity is driven by a coherent state $\beta$
\begin{eqnarray}
U=D_b[\beta].
\end{eqnarray}
Then, the Mollow transformation yields equivalent dynamics as Eqs. \ref{eq:langevin}-\ref{eq:inout} except
\begin{subequations}
\begin{align}
&\dot{{a}}\rightarrow\dot{a}+\sqrt{\kappa}\beta(t)\\
&{b}_\text{out}(t)\rightarrow b_\text{out}(t)-\beta(t).
\end{align}
\end{subequations}
Now the laser pulse in mode $b$ imparts energy to the cavity. After interaction, it is found in the output channel ${b}_\text{out}$.
\end{enumerate}

Importantly, it is possible to show that driving the cavity is equivalent to building up a coherent state $\alpha(t)$ in the cavity, which drives the atom through the coupling~$g$. Consider the transformation $a(t)\rightarrow a(t)+\alpha(t)$ applied to Eqs.~\ref{eq:langevin}-\ref{eq:inout}, where $\alpha(t)=-(\beta*f)(t)$ and 
\begin{eqnarray}
f(t)=\sqrt{\kappa}\,\text{e}^{-(\text{i}\omega_0+\kappa/2)t}\Theta(t)
\end{eqnarray}
is the linear filter response of the cavity and $\Theta$ is the Heaviside function. The coherent field $\alpha(t)$ corresponds to a filtered version of the input pulse, i.e. passing through a bare version of the cavity without the atom. Then, the Mollow transformation gives equivalent dynamics as Eqs. \ref{eq:langevin}-\ref{eq:inout} except
\begin{subequations}
\begin{align}
&\dot{{\sigma}}\rightarrow \dot{\sigma}-g\sigma_z\alpha(t)\\
&\dot{{N}}\rightarrow \dot{N}+g\sigma^\dag \alpha(t) + g\alpha^*(t)\sigma\\
&{b}_\text{out}(t)\rightarrow b_\text{out}(t)-\beta(t)-\sqrt{\kappa}\alpha(t).\label{eq:cavitytransfrombout}
\end{align}
\end{subequations}
This has the remarkable consequence that we drove the cavity field, but found a transformation that turned the Langevin equations into those of the atom being driven. This shows how the two driving cases are, in fact, intimately related to each other.

\begin{figure}[b]
  \includegraphics{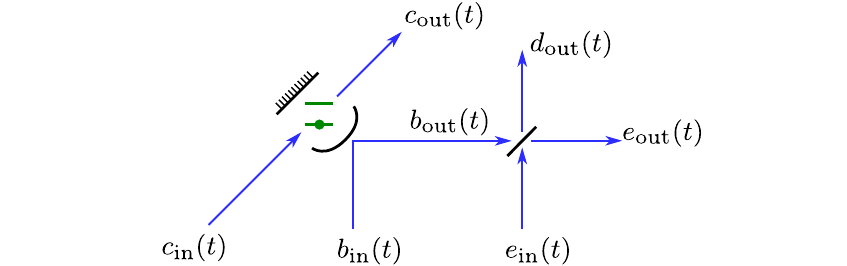}
  \caption{Schematic of the Jaynes--Cummings system with the reservoir $b$ homodyned with field $e$ using a beamsplitter.}
  \label{fig:2}
\end{figure}

Equation \ref{eq:cavitytransfrombout} shows that the output field resulting from driving the cavity is equivalent to directly driving the atom plus an offset of a coherent field $\beta(t)+\sqrt{\kappa}\alpha(t)$. This suggests that by homodyning the output on a beamsplitter with a matching auxiliary field, it is possible to relate the two driving cases perfectly. Hence, imagine we add a beamsplitter (Fig. 3) that mixes the cavity output with a new field $e_\text{in}$. The unitary beamsplitter transformation on the fields is given by
\begin{subequations}\label{eq:BS}
\begin{align}
&e_\text{out}(t)=\frac{1}{\sqrt{2}}\left(-e_\text{in}(t)+b_\text{out}(t)\right) \\
&d_\text{out}(t)=\frac{1}{\sqrt{2}}\left(e_\text{in}(t)+b_\text{out}(t)\right).
\end{align}
\end{subequations}
Additionally, the definition of vacuum is expanded to include the new field.

Now, the cavity is driven by the original coherent state~$\beta$ \textit{and} the beamsplitter channel $e_\text{in}$ is fed a compensatory coherent field $\xi$
\begin{eqnarray}
U=D_b[\beta] D_e[\xi].
\end{eqnarray}
Again, let $\alpha(t)=-(\beta*f)(t)$---then under the replacement $a(t)\rightarrow a(t)+\alpha(t)$, while identifying $\zeta(t)=g\alpha(t)/\sqrt{\gamma}$ and $\xi(t)=\beta(t)+\sqrt{\kappa}\,\alpha(t)$, the Mollow transformation results in equivalent dynamics as Eqs. \ref{eq:langevin}-\ref{eq:inout} and \ref{eq:BS} except
\begin{subequations}
\begin{align}
&\dot{{\sigma}}\rightarrow \dot{\sigma}-\sqrt{\gamma}\sigma_z\zeta(t)\\
&\dot{{N}}\rightarrow \dot{N}+\sqrt{\gamma}\sigma^\dag \zeta(t) + \sqrt{\gamma}\zeta^*(t)\sigma\\
&{b}_\text{out}(t)\rightarrow b_\text{out}(t)-\xi(t)\\
&{d}_\text{out}(t)\rightarrow d_\text{out}(t)-\sqrt{2}\xi(t).
\end{align}
\end{subequations}
Hence, by our appropriate compensation $\xi(t)$, the output field after the beamsplitter $e_\text{out}$ then looks identical to the case where the atom is actually driven by a \textit{real} input field $\zeta(t)$. Here, a compensatory field on a beamsplitter can perfectly cancel the coherent state leaking from the JC system and the output looks as though only the atom were driven!

Because this compensatory field involves the input pulse convolved with the filter response of the bare cavity, $\xi(t)$ could be generated by using a reference cavity (Fig. 4). The pulse $\beta(t)$ would be split between the reference cavity and the Jaynes--Cummings system, and then recombine on a second beamsplitter. A similar configuration, where the cavities are double-sided, has previously been explored for quantum light generation and photonic gates \cite{liew2010single,bamba2011origin,majumdar2012loss,zhang2014optimal,liu2016mode,ye2016simultaneous,wang2017phase}. However, these studies did not identify the root cause of the benefit from the interference, which we explain here as removing an unwanted coherent state that builds up in a Jaynes--Cummings system under cavity drive.

\begin{figure}
  \includegraphics{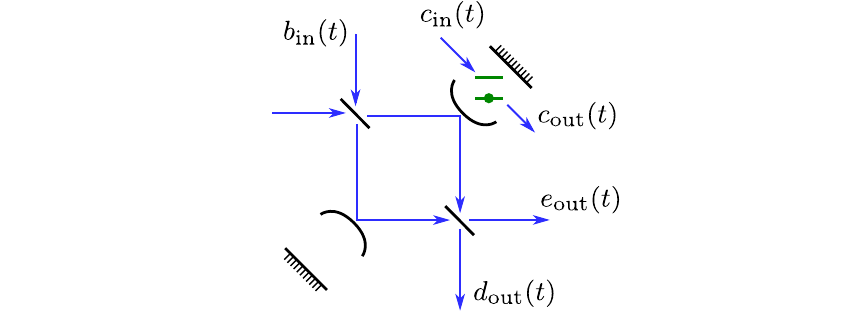}
  \caption{Schematic of a drop filter that automatically generates the appropriate coherent state to compensate the coherent field that builds up in the Jaynes--Cummings system. The first beamsplitter divides the incident pulse and sends it to the JC system and the bare cavity, while the second beamsplitter homodynes the signals together. The response out of the homodyne channel of the drop filter $e_\text{out}$ is the same as if the JC system were instead driven through the atom by the input coherent field convolved with the impulse response of the cavity.}
  \label{fig:3}
\end{figure}

In conclusion, we have theoretically investigated the differences between driving the cavity versus driving the atom of a Jaynes--Cumming system. Traditionally driving the atom has been perceived as providing a much greater nonlinear response from the system. However, we proved that driving the cavity can be thought of as driving the atom through the atom-cavity coupling $g$, with the convolution of the input coherent state with the cavity's linear response. This is important especially for integrated nanophotonic implementations, where driving the atom is much more difficult than driving the cavity. Hence, potentially a drop filter could be used in nanophotonic devices to effectively observe the full nonlinear response from driving the atom \cite{majumdar2012loss}. Alternatively, the homodyning can be performed using a Fano-like structure~\cite{fischer2017chip}, using the full mode structure of a photonic crystal cavity \cite{fischer2016self} or off-chip as well \cite{sun2017cavity}. Finally, we note this type of analysis holds in any situation where an arbitrary nonlinear system is linearly coupled to a set of cavity modes, e.g. when analyzing a cavity linearly coupled to multiple quantum dots \cite{liu2014photon} or a multi-level system \cite{hargart2016probing}.

The authors thank Kai M{\"u}ller for helpful discussions. The authors gratefully acknowledge financial support from the Air Force Office of Scientific Research (AFOSR) MURI Center for Quantum Metaphotonics and Metamaterials and the Army Research Office (ARO) (W911NF1310309 and W911NF1810062). D.L. acknowledges support from the Fong Stanford Graduate Fellowship and the National Defense Science and Engineering Graduate Fellowship. R.T. acknowledges support from the Kailath Stanford Graduate Fellowship.


\bibliography{bibliography}

\end{document}